\documentclass{vgtc}                          

\ifpdf
  \pdfoutput=1\relax                   
  \usepackage{graphicx}                
  \DeclareGraphicsExtensions{.pdf,.png,.jpg,.jpeg} 
\else
  \usepackage{graphicx}                
  \DeclareGraphicsExtensions{.eps}     
\fi%

\graphicspath{{figures/}{pictures/}{images/}{./}} 

\usepackage{microtype}                 
\PassOptionsToPackage{warn}{textcomp}  
\usepackage{textcomp}                  
\usepackage{mathptmx}                  
\usepackage{times}                     
\usepackage{cite}                      
\usepackage{tabu}                      
\usepackage{booktabs}                  

\usepackage{xcolor}

\onlineid{0}

\vgtccategory{Research}

\vgtcinsertpkg

\title{Characterizing Automated Data Insights}

\author{
Po-Ming Law
\thanks{e-mail: pmlaw@gatech.edu}\\ %
\scriptsize Georgia Institute of Technology %
\and 
Alex Endert 
\thanks{e-mail: endert@gatech.edu}\\ %
\scriptsize Georgia Institute of Technology %
\and 
John Stasko
\thanks{e-mail: stasko@cc.gatech.edu}\\ %
\scriptsize Georgia Institute of Technology
}

\abstract{
Many researchers have explored tools that aim to recommend data insights to users. These tools automatically communicate a rich diversity of data insights and offer such insights for many different purposes. However, there is a lack of structured understanding concerning what researchers of these tools mean by “insight” and what tasks in the analysis workflow these tools aim to support. We conducted a systematic review of existing systems that seek to recommend data insights. Grounded in the review, we propose 12 types of automated insights and four purposes of automating insights. We further discuss the design opportunities emerged from our analysis.
} 

\CCScatlist{
  \CCScatTwelve{Human-centered computing}{Visu\-al\-iza\-tion}{Visu\-al\-iza\-tion theory, concepts and paradigms}{}
}

\begin{document}


\firstsection{Introduction}

\maketitle

\begin{figure}[t!]
	\centering
	\includegraphics[width=\linewidth]{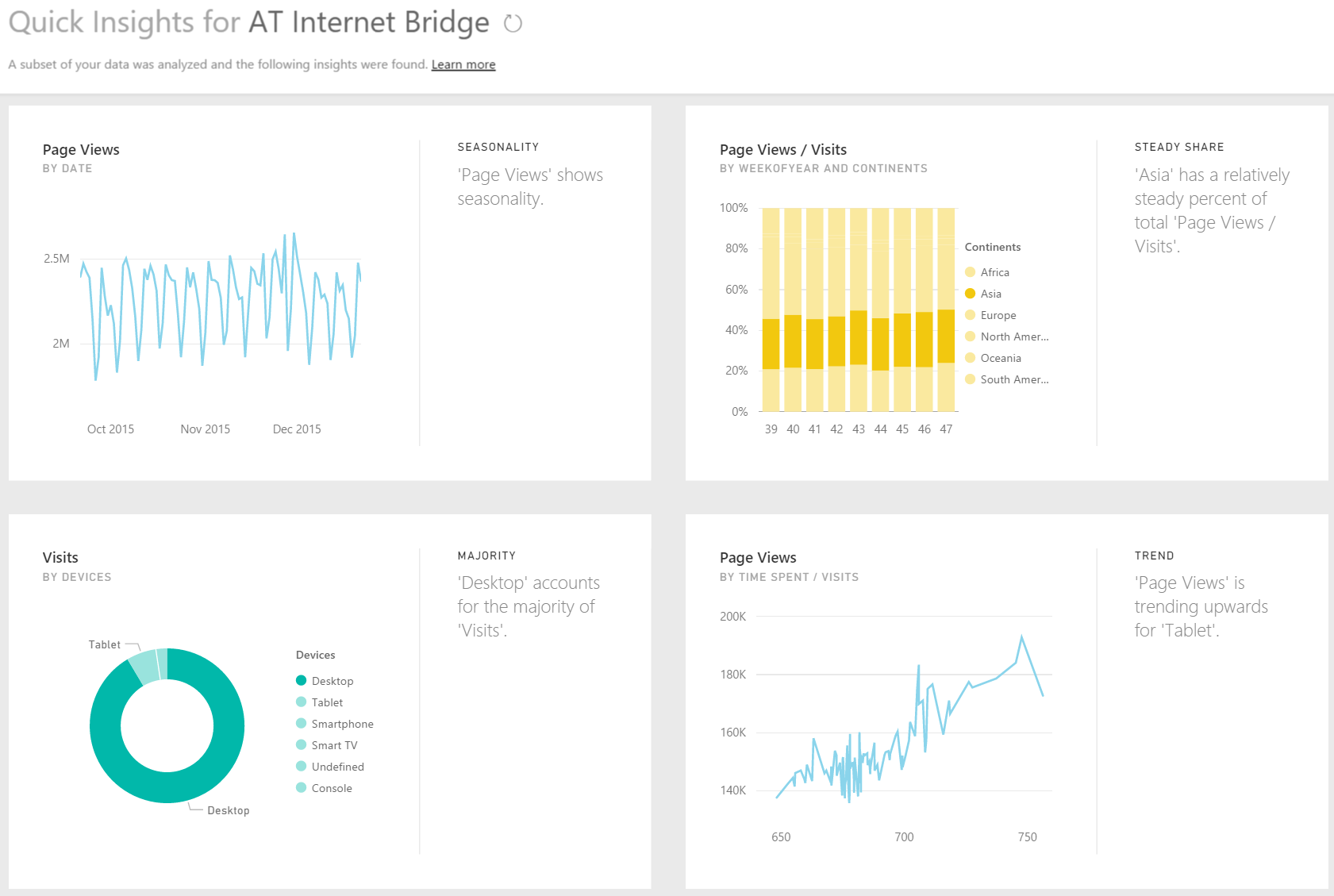}
	\vspace{-5mm}
	\caption{The Quick Insights interface~\cite{quickInsights}. Many visualization systems aim to automatically provide data insights to users.}
	\vspace{-3mm}
	\label{powerBI}
\end{figure}

{


Providing insight has been recognized as a main goal of visualization~\cite{quote}. However, gleaning data insights from visualization is a non-trivial task that requires domain knowledge, analysis expertise, and visualization literary. To facilitate insight generation, some researchers have created systems that automatically communicate data insights to users~\cite{seedb, quickInsights, datashot}. For instance, Quick Insights in Power BI~\cite{quickInsights} suggests prominent trends and patterns within a data set that are presented as charts along with textual descriptions (Fig.~\ref{powerBI}).

Developers of these systems often use the term ``insight'' to refer to the automatically-extracted information (e.g., Quick Insights~\cite{quickInsights} and Automated Insights~\cite{autoInsight}). However, ``insight'' is an overloaded term that has been applied from multiple perspectives in the visualization community~\cite{newPaper}. In the seminal work about insight-based evaluation, Saraiya et al.~\cite{insightEval} regard insights as data findings. On the other hand, Sacha et al.~\cite{knowledge} consider insights a product resulting from evaluating data findings with domain knowledge. Lacking a clarification of what insights are in the context of these automated systems can create confusion to researchers and hinder communication of ideas within the visualization community.

Furthermore, the early development of tools that automate the identification of data insights was motivated by the sentiment that data exploration involves inefficient manual specification of a large number of charts. These early tools (e.g.,~\cite{db1, seedb}) focus on automatically surfacing potentially interesting charts to facilitate exploratory visual analysis. More recently, some researchers have explored the use of such automated tools beyond data exploration. They have investigated new applications such as focused question answering~\cite{me2} and communication~\cite{voder, datashot}. Clearly, we are still working to understand the different purposes of automating data insights.

In order to gain a better understanding of automated systems that suggest data insights, we conducted a systematic review of publications that describe these systems. Based on the review of 20 relevant papers, we propose a framework to organize the \textbf{types} of automatically-generated insights (\textit{auto-insights}) and the \textbf{purposes} of providing auto-insights. We further discuss four design implications based on the analysis. Our work can shed light on the existing landscape of tools that seek to recommend data insights to users.

}

\section{Methodology}

{


While some researchers regard insight as a product of human interpretation of data, in this paper, we define auto-insights as \textit{data observations revealed by automation} (as opposed to human interpretation). This perspective of insight corresponds to Saraiya et al.'s definition of an insight as a unit of data discovery~\cite{insightEval} and the way some laypeople consider data insights~\cite{insightDef}.

We focused on two forms of auto-insights: textual and visual insights. Textual insights (or ``data facts''~\cite{voder, datashot}) are statements that describe statistical facts about data items, subsets, or aggregations (e.g., US cars have a higher average horsepower than Japanese cars for a car dataset). Visual insights are potentially interesting charts (e.g., a scatterplot that shows a high correlation). We call systems that recommend textual or visual insights auto-insight tools.

}

\subsection{Collecting Relevant Papers}

One of our objectives was to understand the purposes of automating insights: What tasks within the data analysis workflow do system designers think auto-insight tools are useful for? Hence, we focused on auto-insight tools that have a graphical user interface. Researchers of these tools often specify a target user group and user tasks. Reviewing related papers helped expose the purposes of automating insights in terms of the tasks auto-insights could facilitate. We excluded papers that describe algorithms for extracting and ranking auto-insights (e.g.,~\cite{algo1}) because they often have foci unrelated to user tasks (e.g., optimizing computational efficiency). We also excluded commercial auto-insight tools that have not been published in academic conferences and journals as they tend to provide fewer details about the auto-insights they support. Our review further centered on tabular data as it is one of the most common data types.

With the focus on auto-insight user interfaces for tabular data, we collected a set of seed papers. We considered both systems that recommend textual insights and those that recommend visual insights. For systems that recommend textual insights, we started with two recent publications that used the term “data facts” to describe the textual insights offered by the systems~\cite{voder, datashot}. For systems that recommend visual insights, Lee~\cite{doris} reviewed 12 relevant auto-insight tools. We found that some (e.g., SeeDB~\cite{seedb}) have multiple publications and included only the most highly cited paper for each tool. From the resulting 12 papers, we omitted two that do not depict user interfaces~\cite{partition, olap}. Hence, we started with 2 (textual insight systems) $+$ 10 (visual insight systems) $=$ 12 seed papers.

We then gathered papers citing the seed papers from Google Scholar and those cited by the seed papers. We found 833 unique papers. Having collected the set, we reviewed only publications at nine relevant venues (InfoVis, VAST, TVCG, EuroVis, SIGCHI, AVI, VLDB, SIGKDD, and SIGMOD). The review resulted in eight additional papers that depict auto-insight tools~\cite{quickInsights, feedback, tsi, profiler, me1, me2, lensxplain, d2insight}. We again collected papers citing the eight papers and those cited by the eight papers and found 298 unique papers. We did not discover additional relevant papers from the set. The paper collection process yielded 12 (original seed papers) $+$ 8 (additional papers) $=$ 20 relevant papers. It occurred in March 2020.

During the process, we omitted some tools that offer recommendations other than textual and visual insights. For instance, Voyager~\cite{voyager1} and Show Me~\cite{showMe} recommend perceptually-effective visualizations but do not proactively identify potentially interesting visualizations based on the statistical properties of data.

\subsection{Coding the Types of Auto-Insights}

We analyzed the 20 papers by coding the types of auto-insights the tools present and the purposes of providing auto-insights to users. 

For the types of auto-insights, we used the fact taxonomy proposed by Chen et al.~\cite{insightMgmt} as a foundation for the analysis. Their taxonomy depicts a comprehensive set of facts that can be discovered from tabular data and matches our definition of auto-insights as statistical facts in visual and/or textual forms. Grounded in our review of relevant papers, we found 12 auto-insight types (e.g., outliers and association). The 12 types comprise 11 types adapted from Chen et al.’s taxonomy~\cite{insightMgmt} and a new type (i.e. visual motifs).

We observed that while some papers clearly depict the types of auto-insights their tools support (e.g.,~\cite{datasite, voder}), others tend to be vague in the description. For example, some tools provide a flexible framework for including a rich set of auto-insights. Yet, the papers do not state clearly which types were included in the implementation. While coding auto-insight types, we looked for explicit statements about the provision of an auto-insight type in the paper and examined the accompanying figures and videos.

\subsection{Coding the Purposes of Automating Insights}

Regarding the purposes of automating insights, we referred to process models that illustrate different stages within the data analysis workflow~\cite{marti, kandel}. Alspaugh et al.~\cite{marti} interviewed professional data analysts and proposed six phases in the analysis process: discover, wrangle, profile, model, explore, and report. Using these phases as a starting point, the coding resulted in four tasks within the data analysis workflow that auto-insight tools intend to support: exploratory analysis, communication, focused analysis, and data wrangling.

The tasks that an auto-insight tool supports can be ambiguous. For example, as stated in the title of the zenvisage paper (``Effortless Data Exploration with zenvisage''), zenvisage intends to support exploratory analysis~\cite{zenvisage}. However, it allows users to draw a line chart to query for charts with a similar temporal trend (focused analysis). Furthermore, an analyst could present the charts recommended by the tool in front of a manager (communication). In order not to over-interpret the tasks that a tool supports, we only coded the \textit{major} purpose that a tool provides auto-insights by reading the title and the introduction in the paper. We coded zenvisage as an exploratory analysis tool because it is emphasized in the title of the paper~\cite{zenvisage}.

\begin{figure}[t!]
	\centering
	\includegraphics[width=\linewidth]{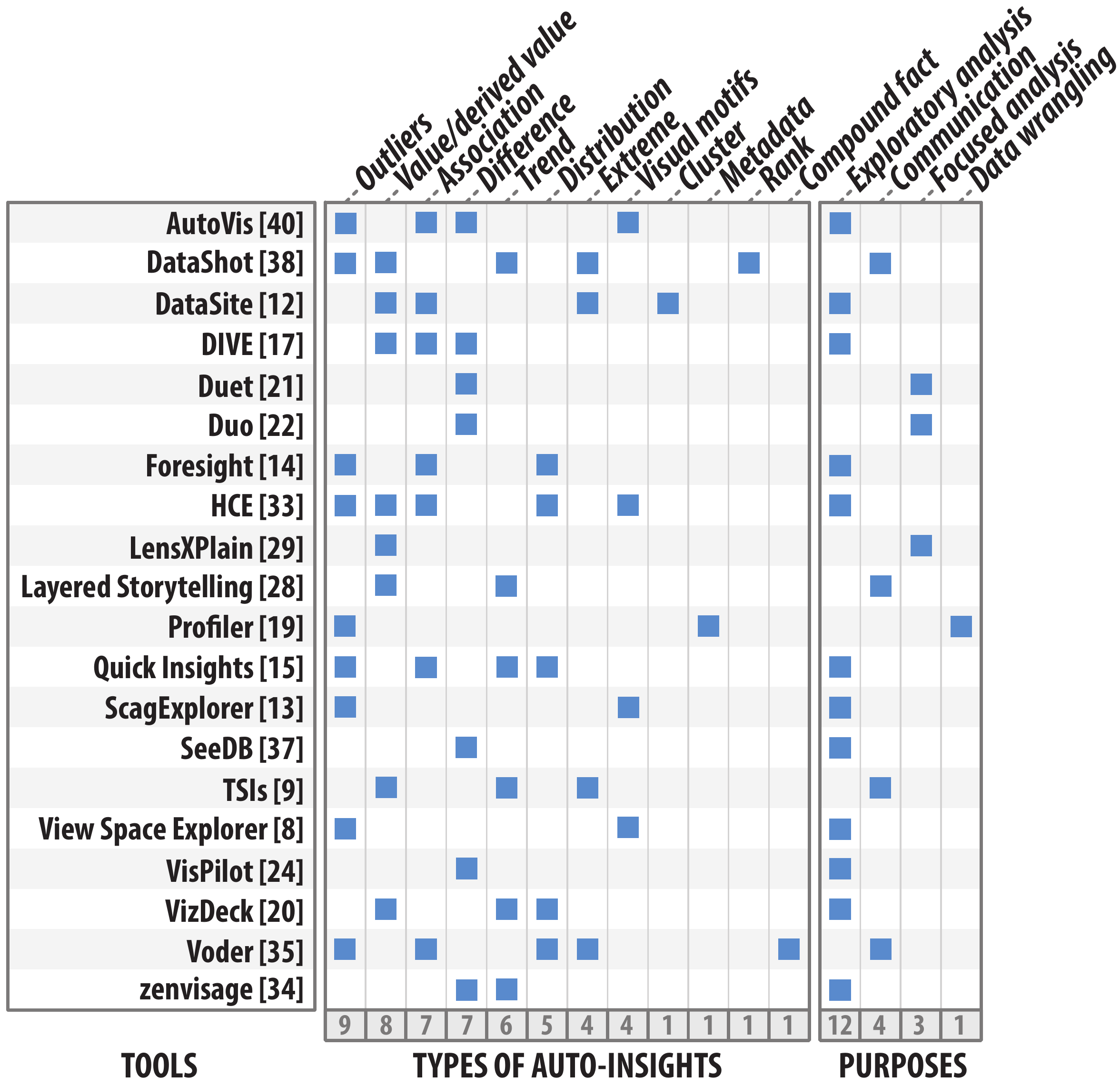}
	\vspace{-6mm}
	\caption{Result of the literature review. Each row is an auto-insight tool. A blue square indicates that a tool provides a type of auto-insights or puts emphasis on a particular purpose of automating insights.}
	\vspace{-3mm}
	\label{table}
\end{figure}

\section{Organizational Framework}

The literature review resulted in 12 types of auto-insights and four purposes of providing auto-insights (Fig.~\ref{table}).

\subsection{Types of Automated Insights}

We observed various approaches to recommending auto-insights. Tools that provide visual insights often score and rank charts to prioritize the presentation of the charts~\cite{general}. For example, to recommend correlation insights, Foresight ranks scatterplots by correlation coefficient~\cite{foresight}. Tools that recommend textual insights may present facts based on the attributes displayed in a chart or the noteworthy regions in a visualization. For instance, Voder generates textual descriptions in relation to the attribute combination in a chart~\cite{voder}. Temporal summary images identify salient regions in a visualization and annotate them with descriptive facts~\cite{tsi}. Below, we present auto-insight types in descending order of occurrence frequency.

\vspace{1mm} \noindent \textbf{Outliers}. Auto-insights about outliers are available in nine auto-insight tools. 5/20 tools provide auto-insights about outliers in a single variable. For example, Voder categorizes data values that are 1.5 times the interquartile range below the first quartile or above the third quartile as outliers~\cite{voder}. 5/20 tools provide auto-insights about outliers in two dimensions. Notable examples are the systems that employ scagnostics for ranking scatterplots (e.g.,~\cite{feedback,db1}). These systems utilize the outlying scagnostics measure to identify scatterplots with outlying points. Finally, Quick Insights communicates outliers in a time series that are highlighted in a line chart~\cite{quickInsights}.

\vspace{1mm} \noindent \textbf{Value/derived value}. Auto-insights about a value of a single row in a table or a value derived from multiple rows in the table appear in 8/20 tools. The multimodal layered storytelling approach reveals prominent values in a timeline~\cite{d2insight}. VizDeck measures the number of unique categories in a categorical variable and ranks bar charts by the unique category count~\cite{vizdeck}. DataSite finds the average of a numerical variable~\cite{datasite} while DataShot computes the proportion of categories in a categorical variable~\cite{datashot}. Both systems state the derived values (average and proportion) as textual descriptions.

\vspace{1mm} \noindent \textbf{Association}. 7/20 tools provide auto-insights about association (i.e. quantitative relationship between two numerical variables). These tools commonly identify either linear relationship using Pearson correlation (7/20) or non-linear relationship using more sophisticated measures (1/20). DataSite computes a Pearson correlation between two variables and presents it as a textual description alongside a scatterplot~\cite{datasite}. Hierarchical Clustering Explorer (HCE) ranks scatterplots by least-squares error curvilinear regression and quadracity to identify scatterplots that show a quadratic relationship~\cite{rank}. 

\vspace{1mm} \noindent \textbf{Difference}. An auto-insight about difference involves a quantitative comparison between distributions. 7/20 tools provide such auto-insights. With Duo~\cite{me2}, users can specify two groups of objects (e.g., cities in China and cities in the US). For each attribute (e.g., population), Duo compares the two groups to determine whether they have different distributions~\cite{me2}. SeeDB ranks grouped bar charts by computing the earth mover’s distance between the two probability distributions depicted in the charts~\cite{seedb}. AutoVis conducts a one-way ANOVA for charts that show continuous-categorical data and sorts the charts by p-value~\cite{autovis}.

\vspace{1mm} \noindent \textbf{Trend}. We found auto-insights about temporal trend in 6/20 tools. These tools extract upward and downward trends (3/20), steady trend (2/20), and periodicity (2/20). With the ZQL language for specifying visualizations, zenvisage can order line charts based on the upward trends they show~\cite{zenvisage}. Temporal summary images add annotations to the flat regions in a time series visualization~\cite{tsi}. Quick Insights extracts time series that show seasonality~\cite{quickInsights}.

\vspace{1mm} \noindent \textbf{Distribution}. 5/20 systems communicate auto-insights about the distribution of a variable. Voder presents data facts about the range of a numerical variable~\cite{voder}. Foresight ranks charts based on several measures of distribution including dispersion, skewness, and heavy-tailedness to reveal charts with a noteworthy distribution~\cite{foresight}. 

\vspace{1mm} \noindent \textbf{Extreme}. 4/20 tools show auto-insights about the minimum and maximum values in a stream of values. For example, DataSite~\cite{datasite} and Voder~\cite{voder} present the minimum and maximum values in a numerical variable as textual descriptions. Temporal summary images annotate the lowest and highest points in a time series~\cite{tsi}. 

{ \vspace{1mm} \noindent \textbf{Visual motifs}. Visual motifs are unique patterns in a chart that do not fall into other auto-insight types.} They include the special patterns in scatterplots identified by the scagnostics measures~\cite{scag}. For example, the striated measure finds scatterplots with parallel lines. 4/20 tools identify visual motifs in scatterplots by utilizing scagnostics (~\cite{autovis, db1, feedback}) or other measures (e.g., uniformity)~\cite{rank}.

\vspace{1mm} \noindent \textbf{Cluster}. Only DataSite recommends auto-insights about clusters. DataSite employs K-means and DBSCAN to find clusters in a scatterplot~\cite{datasite}. It presents the clusters using a textual description and highlights the clusters in the scatterplot~\cite{datasite}.

\vspace{1mm} \noindent \textbf{Metadata}. Auto-insights about metadata provide information about a dataset. Such information includes missing values and other data quality issues~\cite{insightMgmt}. Profiler uses detection routines to identify data quality issues and suggests charts to visualize the issues~\cite{profiler}. 

\vspace{1mm} \noindent \textbf{Rank}. Such auto-insights involve sorting categories by a numerical variable. For a dataset of cars, DataShot recommends a data fact that says, ``Compact, SUV, Midsize are the top 3 categories in the year of 2008''~\cite{voder}. This fact is generated by ranking different types of cars by the numerical variable sales.

\vspace{1mm} \noindent \textbf{Compound fact}. Chen et al.~\cite{insightMgmt} defined a compound fact as “a fact that contains two or more facts.” Voder recommends a fact by combining auto-insights about derived value and distribution~\cite{voder}. For example, it generates the fact ``Average Retail Price of SUV is 1.76 times Sedan'' for a car dataset~\cite{voder}. The fact includes a derived value (average) and is about the distribution of retail price.

\subsection{Purposes of Automating Insights}

While the previous section regards what types of auto-insights the reviewed tools provide, this section illuminates how these auto-insights support tasks within the data analysis workflow. Early research effort on auto-insight tools centered on supporting open-ended data exploration. However, we observed new applications of such tools in other aspects of data analysis.

\vspace{1mm} \noindent \textbf{Exploratory analysis}. The majority of the auto-insight tools we reviewed intend to support exploratory data analysis (12/20). Several papers argue that data exploration typically involves specifying and examining a large number of charts (5/20). They comment that the process is non-trivial because of a large dataset (3/20) or a limited expertise of users (4/20). Tools such as Quick Insights~\cite{quickInsights} and Foresight~\cite{foresight} therefore rank charts based on the statistical properties of data (e.g., correlation) and show the potentially interesting ones to reduce the number of charts for user review.

Aside from surfacing potentially interesting charts, we observed other reasons for providing auto-insights to support data exploration. Will and Wilkinson~\cite{autovis} argue that analysts might not know where to look at the beginning of data exploration, and automatically revealing interesting charts helps analysts enter the exploratory loop. Seo and Shneiderman~\cite{rank} feel that visualization systems often leave analysts ``uncertain about how to explore their data in an orderly manner.'' They created the HCE system to support a more systematic data exploration~\cite{rank}. Lee et al.~\cite{pilot} suggest that users often encounter drill-down fallacies and developed VisPilot that recommends bar charts to protect users from the analysis pitfall.

\vspace{1mm} \noindent \textbf{Communication}. More recently, some researchers have developed tools that automatically generate data insights for communication purposes (4/20). These tools commonly present textual insights beside visualizations. They utilize textual insights to provide various benefits during communication with visualizations, including visualization interpretation, effective storytelling, and reflection.

Voder seeks to scaffold visualization interpretation by presenting textual descriptions of charts~\cite{voder}. DataShot~\cite{datashot} and Temporal summary images~\cite{tsi} focus on data-driven storytelling. DataShot generates infographic-like fact sheets to communicate key points in a dataset~\cite{datashot}. Temporal summary images annotate temporal visualizations to point ``a viewer’s attention to regions of interest,'' ``suggest conclusions,'' and ``provide data context''~\cite{tsi}. Martinez-Maldonado et al.~\cite{d2insight} further found that automatically generated annotations of visualizations served educational purposes by encouraging reflection on performance in nursing simulations.

\vspace{1mm} \noindent \textbf{Focused analysis}. Analysis exists on a spectrum from exploratory to directed. Data exploration is often more opportunistic and involves a vague goal while focused analysis is directed by more concrete questions. We found 3/20 auto-insight tools that were designed to support more focused analysis. They help answer concrete yet high-level questions such as ``Why is there a large high-income White population?''~\cite{lensxplain} and ``What are the differences between New England colleges and Southeast colleges?''~\cite{me2} While low-level questions such as “What is the admission rate of University X in 2020?” often have a single correct answer, high-level questions may have multiple reasonable answers. Auto-insight tools lend themselves to answering high-level questions by recommending possible answers. Users can then apply their domain knowledge to interpret the relevance of the recommendations.

\vspace{1mm} \noindent \textbf{Data wrangling}. Another purpose of automatically extracting data insights is to support data wrangling. In the set of tools we reviewed, only Profiler serves this purpose. It detects anomalies in data and recommends visualizations to show the data quality issues~\cite{profiler}.

\section{Discussion}

Here, we discuss the design opportunities that emerged during the literature review and reflect on the limitations of our work.

\subsection{Design Opportunities}

Our organizational framework enables comparison of existing tools that aim to automate data insights. Reviewing the literature using our framework highlights prevailing approaches to automating insights. Furthermore, our review helps inspire new approaches by identifying under-examined spaces. This section discusses four design opportunities based on the observations from our analysis.

\vspace{1mm} \noindent \textbf{Compound facts}. Figure~\ref{table} indicates several types of auto-insights that are rarely provided. A promising research avenue is the provision of compound facts. The auto-insight tools we reviewed often provide relatively simple facts. For a car data set, these facts might suggest whether a correlation is high or low or whether a temporal trend is upward or downward. While Voder generates compound facts, in the current implementation, these facts (e.g., “Average Retail Price of SUV is 1.76 times Sedan”) appear to be straightforward~\cite{voder}. Some visualization researchers feel that the auto-insights generated by existing tools do not align with the conceptualization of human insights being deep and complex~\cite{stasko}. The view that existing auto-insights lack depth imply opportunities for investigating the feasibility and utility of generating more nuanced auto-insights that involve multiple auto-insight types and more sophisticated logic. 

However, communicating more nuanced insights may involve a different set of design considerations. While text lends itself to communicating multiple pieces of information, creating effective visualizations to highlight different information at once is non-trivial. If designed properly, however, visualizations can ease user effort in interpreting the information. Developing algorithms for extracting compound facts and investigating the appropriate presentations of these facts are ripe for future research.

\vspace{1mm} \noindent \textbf{Beyond exploratory analysis}. While many researchers have designed auto-insight tools for exploratory analysis, new applications have emerged. One such application is focused analysis. Tableau Explain Data~\cite{explainData} offers a recent example. As users select an outlying value, Explain Data automatically generates potential explanations for why the value is unusually high or low. However, many questions concerning focused analysis remain to be addressed. Future work will investigate other common high-level questions that auto-insight tools can help answer. In designing these tools, researchers should also investigate how these automated systems affect users during data analysis. Do they really make us a better analyst? Do they introduce any side effects to the analysis process?

Besides the current applications of auto-insight tools, research could be devoted to exploring new application areas. We note that suggesting new applications entails conducting studies with real users to understand their needs. During the literature review, we noticed that auto-insight tools informed by formative user studies in specific domains are scarce.  Exceptions include a tool created by Martinez-Maldonado et al.~\cite{d2insight} who found that providing textual insights alongside a timeline helped support reflection in an education setting. Without significant understanding of users, the new applications identified may be divorced from real-world needs.

\vspace{1mm} \noindent \textbf{Explanations}. Some researchers were concerned that a lack of transparency in auto-insight tools will cause user distrust of these tools~\cite{doris}. Lee~\cite{doris} provides an example of the non-transparency: How do users know that the insights provided ``cover all the things that can be learned from the dataset''? In general, users may want to know why the auto-insights are generated and how they are generated. Although explanations have been recognized as an approach to inspiring user trust and promoting transparency in automated systems~\cite{mythos}, none of the tools we reviewed explicitly provides explanations about the generation of auto-insights.

Lessons could be learned from other research areas regarding the provision of explanations. In context-aware systems, recommender systems, and machine learning systems, much research has been devoted to investigating the types of explanations that can be provided and the effectiveness of providing these explanations. For instance, Lim et al.~\cite{lim} proposed five types of explanations for context-aware systems (i.e. what, why, why not, what if, and how to). Herlocker et al.~\cite{herlocker} illustrated the process of automated collaborative filtering and derived explanations based on the operations in the process. 

These research areas hint at the types of explanations auto-insight tools can provide to users. Based on Lim et al.’s work~\cite{lim}, an auto-insight tools can provide what, why, why-not, what-if, and how-to explanations. A why explanation describes why an auto-insight is recommended to users while a why-not explanation provides reasons for why an auto-insight is not recommended. 

Grounded in the research by Herlocker et al.~\cite{herlocker}, a tool can explain the auto-insights by revealing the generation process. For instance, to explain an auto-insight about the correlation between two variables, a tool can describe how it extracts the two columns and removes the records that have missing values. It can then explain how it computes a correlation coefficient and shows the auto-insight because the correlation is higher than a threshold. Future work is required to understand the effectiveness of such explanations in improving transparency of auto-insight tools.

\vspace{1mm} \noindent \textbf{Information overload}. Our review identified reducing information load as an important purpose of automating insights. Developers of auto-insight tools often argue that data exploration is non-trivial because users need to examine a large number of charts, and that recommending potentially interesting charts reduces information load~\cite{seedb}. As system developers strive to provide more complex auto-insights and explanations for how and why the auto-insights are generated, information overload might become a concern that defeats the original purpose of surfacing auto-insights. This is challenging because developers will need to strike a balance among providing rich auto-insights, instilling user trust, and avoiding information overload. We hope that our work will inspire researchers to consider a wide range of aspects of auto-insight tools that might affect users in different tasks within the data analysis workflow.

\subsection{Limitations}

To keep the literature review manageable and focused, we only considered a representative set of tools that have a graphical user interface, that mine auto-insights from tabular data, and that appeared in top conferences and journals. Aside from the notable landmark publications we surveyed, future work will examine a broader set of publications that describe algorithms for auto-insight generation, that mine auto-insights from other types of data, and that have not been published (e.g., commercial products). Our work can offer foundational understanding of existing approaches to automating insights for future literature reviews to build on.

{


While the types and purposes of auto-insights constitute two significant design dimensions of auto-insight tools, investigating other dimensions can paint a more holistic picture of the landscape of these tools. Moving forward, future work could explore other technical aspects (e.g., the techniques employed for mining the auto-insights) and design aspects (e.g., the methods for evaluating the data insights offered by the tools) of automating data insights.

}

Finally, our literature review started with a selection of seed papers. The seed papers might bias the final set of auto-insight tools we obtained. The 20 auto-insight tools we found, however, could serve as a resource for other researchers to collect a more comprehensive set of auto-insight tools in future studies.

\section{Conclusion}

In this paper, we have proposed a framework to organize tools that aim to automatically communicate data insights. Grounded in a review of 20 auto-insight tools, we identified 12 types of auto-insights and four purposes of offering these insights. We further discussed four design opportunities that emerged from the review. Resonating with an ongoing research endeavor to understand the automation of data insights, we hope that our work will offer more consolidated understanding of existing tools that seek to recommend data insights to users.

\nocite{dive}
\bibliographystyle{abbrv-doi}

\bibliography{template}

\begin{thebibliography}{10}

\bibitem{autoInsight}
Automated {I}nsights: Natural language generation.
\newblock \\ \url{https://automatedinsights.com}.
\newblock Accessed: 2020-03-31.

\bibitem{explainData}
Explain {D}ata $|$ {T}ableau {S}oftware.
\newblock \url{https://www.tableau.com/products/new-features/explain-data}.
\newblock Accessed: 2020-03-31.

\bibitem{doris}
Insight machines: The past, present, and future of visualization
  recommendation.
\newblock
  \href{https://medium.com/multiple-views-visualization-research-explained/insight-machines-the-past-present-and-future-of-visualization-recommendation-2185c33a09aa}{\texttt{https://medium.com...}}
\newblock Accessed: 2020-03-31.

\bibitem{insightDef}
What are data insights?
\newblock \url{https://algorithmia.com/blog/what-are-data-insights}.
\newblock Accessed: 2020-03-31.

\bibitem{stasko}
What's an insight? – as i see it.
\newblock \url{https://jts3blog.wordpress.com/2018/02/22/whats-an-insight}.
\newblock Accessed: 2020-03-31.

\bibitem{marti}
S.~Alspaugh, N.~Zokaei, A.~Liu, C.~Jin, and M.~A. Hearst.
\newblock Futzing and moseying: Interviews with professional data analysts on
  exploration practices.
\newblock {\em IEEE Transactions on Visualization and Computer Graphics},
  25(1):22--31, 2018.

\bibitem{partition}
A.~Anand and J.~Talbot.
\newblock Automatic selection of partitioning variables for small multiple
  displays.
\newblock {\em IEEE Transactions on Visualization and Computer Graphics},
  22(1):669--677, 2015.

\bibitem{feedback}
M.~Behrisch, F.~Korkmaz, L.~Shao, and T.~Schreck.
\newblock Feedback-driven interactive exploration of large multidimensional
  data supported by visual classifier.
\newblock In {\em 2014 IEEE Conference on Visual Analytics Science and
  Technology (VAST)}, pp. 43--52. IEEE, 2014.

\bibitem{tsi}
C.~Bryan, K.-L. Ma, and J.~Woodring.
\newblock Temporal {S}ummary {I}mages: An approach to narrative visualization
  via interactive annotation generation and placement.
\newblock {\em IEEE Transactions on Visualization and Computer Graphics},
  23(1):511--520, 2016.

\bibitem{quote}
S.~K. Card, J.~D. Mackinlay, and B.~Shneiderman.
\newblock {\em Readings in information visualization: Using vision to think}.
\newblock Morgan Kaufmann, 1999.

\bibitem{insightMgmt}
Y.~Chen, J.~Yang, and W.~Ribarsky.
\newblock Toward effective insight management in visual analytics systems.
\newblock In {\em 2009 IEEE Pacific Visualization Symposium}, pp. 49--56. IEEE,
  2009.

\bibitem{datasite}
Z.~Cui, S.~K. Badam, M.~A. Yal{\c{c}}in, and N.~Elmqvist.
\newblock Data{S}ite: Proactive visual data exploration with computation of
  insight-based recommendations.
\newblock {\em Information Visualization}, 18(2):251--267, 2019.

\bibitem{db1}
T.~N. Dang and L.~Wilkinson.
\newblock Scag{E}xplorer: Exploring scatterplots by their scagnostics.
\newblock In {\em 2014 IEEE Pacific Visualization Symposium}, pp. 73--80. IEEE,
  2014.

\bibitem{foresight}
{\c{C}}.~Demiralp, P.~J. Haas, S.~Parthasarathy, and T.~Pedapati.
\newblock Foresight: Recommending visual insights.
\newblock {\em Proceedings of the VLDB Endowment}, 10(12):1937--1940, 2017.

\bibitem{quickInsights}
R.~Ding, S.~Han, Y.~Xu, H.~Zhang, and D.~Zhang.
\newblock Quick{I}nsights: Quick and automatic discovery of insights from
  multi-dimensional data.
\newblock In {\em Proceedings of the 2019 International Conference on
  Management of Data}, pp. 317--332, 2019.

\bibitem{herlocker}
J.~L. Herlocker, J.~A. Konstan, and J.~Riedl.
\newblock Explaining collaborative filtering recommendations.
\newblock In {\em Proceedings of the 2000 ACM conference on Computer Supported
  Cooperative Work}, pp. 241--250, 2000.

\bibitem{dive}
K.~Hu, D.~Orghian, and C.~Hidalgo.
\newblock D{IVE}: A mixed-initiative system supporting integrated data
  exploration workflows.
\newblock In {\em Proceedings of the Workshop on Human-In-the-Loop Data
  Analytics}, pp. 1--7, 2018.

\bibitem{kandel}
S.~Kandel, A.~Paepcke, J.~M. Hellerstein, and J.~Heer.
\newblock Enterprise data analysis and visualization: An interview study.
\newblock {\em IEEE Transactions on Visualization and Computer Graphics},
  18(12):2917--2926, 2012.

\bibitem{profiler}
S.~Kandel, R.~Parikh, A.~Paepcke, J.~M. Hellerstein, and J.~Heer.
\newblock Profiler: Integrated statistical analysis and visualization for data
  quality assessment.
\newblock In {\em Proceedings of the International Working Conference on
  Advanced Visual Interfaces}, pp. 547--554, 2012.

\bibitem{vizdeck}
A.~Key, B.~Howe, D.~Perry, and C.~Aragon.
\newblock Viz{D}eck: self-organizing dashboards for visual analytics.
\newblock In {\em Proceedings of the 2012 International Conference on
  Management of Data}, pp. 681--684, 2012.

\bibitem{me1}
P.-M. Law, R.~C. Basole, and Y.~Wu.
\newblock Duet: Helping data analysis novices conduct pairwise comparisons by
  minimal specification.
\newblock {\em IEEE Transactions on Visualization and Computer Graphics},
  25(1):427--437, 2018.

\bibitem{me2}
P.-M. Law, S.~Das, and R.~C. Basole.
\newblock Comparing apples and oranges: Taxonomy and design of pairwise
  comparisons within tabular data.
\newblock In {\em Proceedings of the SIGCHI Conference on Human Factors in
  Computing Systems}, pp. 1--12, 2019.

\bibitem{newPaper}
P.-M. Law, A.~Endert, and J.~Stasko.
\newblock What are data insights to professional visualization users?
\newblock In {\em IEEE Visualization Conference (VIS)}. IEEE, 2020.

\bibitem{pilot}
D.~J.-L. Lee, H.~Dev, H.~Hu, H.~Elmeleegy, and A.~Parameswaran.
\newblock Avoiding drill-down fallacies with {V}is{P}ilot: Assisted exploration
  of data subsets.
\newblock In {\em Proceedings of the 24th International Conference on
  Intelligent User Interfaces}, pp. 186--196, 2019.

\bibitem{lim}
B.~Y. Lim, A.~K. Dey, and D.~Avrahami.
\newblock Why and why not explanations improve the intelligibility of
  context-aware intelligent systems.
\newblock In {\em Proceedings of the SIGCHI Conference on Human Factors in
  Computing Systems}, pp. 2119--2128, 2009.

\bibitem{mythos}
Z.~C. Lipton.
\newblock The mythos of model interpretability.
\newblock {\em Queue}, 16(3):31--57, 2018.

\bibitem{showMe}
J.~Mackinlay, P.~Hanrahan, and C.~Stolte.
\newblock Show me: Automatic presentation for visual analysis.
\newblock {\em IEEE Transactions on Visualization and Computer Graphics},
  13(6):1137--1144, 2007.

\bibitem{d2insight}
R.~Martinez-Maldonado, V.~Echeverria, G.~F. Nieto, and S.~B. Shum.
\newblock From data to insights: A layered storytelling approach for multimodal
  learning analytics.
\newblock 2020.

\bibitem{lensxplain}
Z.~Miao, A.~Lee, and S.~Roy.
\newblock Lens{XP}lain: Visualizing and explaining contributing subsets for
  aggregate query answers.
\newblock {\em Proceedings of the VLDB Endowment}, 12(12):1898--1901, 2019.

\bibitem{knowledge}
D.~Sacha, A.~Stoffel, F.~Stoffel, B.~C. Kwon, G.~Ellis, and D.~A. Keim.
\newblock Knowledge generation model for visual analytics.
\newblock {\em IEEE Transactions on Visualization and Computer Graphics},
  20(12):1604--1613, 2014.

\bibitem{insightEval}
P.~Saraiya, C.~North, and K.~Duca.
\newblock An insight-based methodology for evaluating bioinformatics
  visualizations.
\newblock {\em IEEE Transactions on Visualization and Computer Graphics},
  11(4):443--456, 2005.

\bibitem{olap}
S.~Sarawagi, R.~Agrawal, and N.~Megiddo.
\newblock Discovery-driven exploration of {OLAP} data cubes.
\newblock In {\em International Conference on Extending Database Technology},
  pp. 168--182. Springer, 1998.

\bibitem{rank}
J.~Seo and B.~Shneiderman.
\newblock A rank-by-feature framework for interactive exploration of
  multidimensional data.
\newblock {\em Information Visualization}, 4(2):96--113, 2005.

\bibitem{zenvisage}
T.~Siddiqui, A.~Kim, J.~Lee, K.~Karahalios, and A.~Parameswaran.
\newblock Effortless data exploration with zenvisage: An expressive and
  interactive visual analytics system.
\newblock {\em Proceedings of the VLDB Endowment}, 10(4), 2016.

\bibitem{voder}
A.~Srinivasan, S.~M. Drucker, A.~Endert, and J.~Stasko.
\newblock Augmenting visualizations with interactive data facts to facilitate
  interpretation and communication.
\newblock {\em IEEE Transactions on Visualization and Computer Graphics},
  25(1):672--681, 2018.

\bibitem{algo1}
B.~Tang, S.~Han, M.~L. Yiu, R.~Ding, and D.~Zhang.
\newblock Extracting top-k insights from multi-dimensional data.
\newblock In {\em Proceedings of the 2017 International Conference on
  Management of Data}, pp. 1509--1524, 2017.

\bibitem{seedb}
M.~Vartak, S.~Rahman, S.~Madden, A.~Parameswaran, and N.~Polyzotis.
\newblock See{DB}: Efficient data-driven visualization recommendations to
  support visual analytics.
\newblock {\em Proceedings of the VLDB Endowment}, 8(13):2182--2193, 2015.

\bibitem{datashot}
Y.~Wang, Z.~Sun, H.~Zhang, W.~Cui, K.~Xu, X.~Ma, and D.~Zhang.
\newblock Data{S}hot: Automatic generation of fact sheets from tabular data.
\newblock {\em IEEE Transactions on Visualization and Computer Graphics},
  26(1):895--905, 2019.

\bibitem{scag}
L.~Wilkinson, A.~Anand, and R.~Grossman.
\newblock Graph-theoretic scagnostics.
\newblock In {\em IEEE Symposium on Information Visualization, 2005. INFOVIS
  2005.}, pp. 157--164. IEEE, 2005.

\bibitem{autovis}
G.~Wills and L.~Wilkinson.
\newblock Auto{V}is: Automatic visualization.
\newblock {\em Information Visualization}, 9(1):47--69, 2010.

\bibitem{voyager1}
K.~Wongsuphasawat, D.~Moritz, A.~Anand, J.~Mackinlay, B.~Howe, and J.~Heer.
\newblock Voyager: Exploratory analysis via faceted browsing of visualization
  recommendations.
\newblock {\em IEEE Transactions on Visualization and Computer Graphics},
  22(1):649--658, 2015.

\bibitem{general}
K.~Wongsuphasawat, D.~Moritz, A.~Anand, J.~Mackinlay, B.~Howe, and J.~Heer.
\newblock Towards a general-purpose query language for visualization
  recommendation.
\newblock In {\em Proceedings of the Workshop on Human-In-the-Loop Data
  Analytics}, pp. 1--6, 2016.

\end{thebibliography}
\end{document}